\documentclass[letterpaper]{aa}
\usepackage[dvips]{graphicx}
\usepackage{txfonts}
\usepackage{natbib}
\bibpunct{(}{)}{;}{a}{}{,} % to follow the A&A style
\begin{document}

\title{Multiple-Images in the Cluster Lens Abell 2218:\\
Constraining the Geometry of the Universe ?}

\author{G. Soucail \inst{1}
          \and
	  J.-P. Kneib\inst{1,2} 
	  \and 
          G. Golse \inst{1} 
          }

\offprints{Genevi\`eve Soucail, \email{soucail@ast.obs-mip.fr}}
\institute{Laboratoire d'Astrophysique, UMR5572 du CNRS,
       Observatoire Midi-Pyr\'en\'ees,
       14 Avenue Belin, F-31400 Toulouse, France
         \and
	California Institute of Technology,  MC105-24, Pasadena, CA 91125, USA
             }

   \date{Received / Accepted }

\abstract{
In this {\it Letter} we present a detailed study of the lensing
configuration in the cluster Abell 2218. Four multiple-images systems
with measured spectroscopic redshifts have been identified in this
cluster. These multiple images are very useful to constrain accurately
the mass distribution in the cluster core, but they are also sensitive
to the value of the geometrical cosmological parameters of the
Universe.  Using a simplified maximum likelihood analysis we find
$0<\Omega_{\rm M}<0.30$ assuming a flat Universe, and $0<\Omega_{\rm
M}<0.33$ and $w<-\, 0.85$ for a flat Universe with dark
energy. Interestingly, an Einstein-de Sitter model is excluded at more
than 4$\sigma$. These constraints are consistent with the current
constraints derived with CMB anisotropies or supernovae studies. The
proposed method constitutes an independent test of the geometrical
cosmological parameters of the Universe and we discuss the limits of
this method and this particular application to Abell 2218.
Application of this method with more sophisticated tools and to a
larger number of clusters or with more multiple images constraints,
will put stringent constraints on the geometrical cosmological
parameters.

   }
\titlerunning{Abell 2218: Constraining the geometry of the Universe?}

   \maketitle
%
%________________________________________________________________

\section{Introduction}

The present Cosmology framework is characterized by a number of
parameters which sets the global geometry of the Universe, its history
and dynamics. The quest for these parameters is a long-standing issue
in Observational Cosmology and is still the main driver of a large
number of experiments.  Combining constraints coming from the power
spectrum of the CMB anisotropies and the luminosity distances of
distant type Ia supernovae (SNIa), a new standard model of cosmology
is emerging \citep{spergel03}: a flat Universe with an accelerating
expansion ($\Omega_{\rm M} \simeq 0.27$ and $\Omega_\Lambda \simeq
0.73$).  To quantitatively explain these results the concept of dark
energy has been put forward, characterized by the ratio of pressure
and energy density $w=P_{\rm X}/\rho_{\rm X} \, c^2$, which reduces to
the vacuum energy (the cosmological constant) for $w=-1$.  There is
however, no strong observational constraints on $w$ yet
(\citealt{spergel03} give only $w<-\, 0.78$).

This new standard cosmology is getting very popular. Although the
flatness of the Universe seems robust, the exact value of $\Omega_{\rm
M}$ is still a matter of debate \citep{bridle03,blanchard03} as it is
essentially driven by the SNIa results which can be discussed
\citep{rowan02}.  In order to independently probe the large scale
geometry of the Universe, we propose to explore the potential use of
cluster lenses as a long range optical test bench. Preliminary
analysis of this method was first detailed by
\citet{link98} using simple lens models. Recently, we extended
their work using more detailed simulations of realistic clusters of
galaxies \citep{golse02a}. The basic idea of this method is that each
set of multiple-images identified in a cluster lens strongly
constrains the cluster potential.  As the scaling of the mass model
depends on the ratio of the angular distances $D_{\rm LS}/D_{\rm OS}$,
it will also depends on the geometrical cosmological parameters
($\Omega_{\rm M}$, $\Omega_\Lambda$ and $w$).  In order to constrain these
parameters, the combination of several sets of multiple images in a
single lens is mandatory to disentangle between the degeneracies in
the lens model. With a minimum of 4 systems of multiple images with
known spectroscopic redshifts, we showed \citep{golse02a} that one can
put reasonable constraints in the ($\Omega_{\rm M} ,
\Omega_\Lambda$) plane with some characteristic degeneracies
in the fitted parameters \citep{golse02a}.

In this {\it Letter} we apply this lensing test to the well studied
cluster-lens Abell 2218.  Section 2 describes details of the lens
modeling and the cluster mass distribution. The results of the
optimization are discussed in Section 3 and a conclusion is presented
in Section 4.  When necessary we scale the physical parameters with
$\Omega_{\rm M} = 0.3, \ \Omega_\Lambda = 0.7, \ h=0.65$ with Hubble
constant $H_0 = 100 \ h$ km s$^{-1}$ Mpc$^{-1}$. Thus at the cluster
redshift $z=0.176$, $1''$ corresponds to a linear scale of
$2.08\,h^{-1}$ kpc.

\section{Lens Modeling}
\subsection{Lensing and other constraints}
Abell\,2218 is one of the richest clusters in the Abell catalog 
\citep{abell89}. A spectroscopic survey of the galaxies by
\citet{leborgne92} led to an average redshift $z_{\rm L}=0.1756$ and a
galaxy velocity dispersion $\sigma=1370^{+160}_{-120}\,$km\,s$^{-1}$
(based on 50 cluster members). It is also one of the few clusters for
which both an accurate lens model and the identification of 4 systems
of lensed multiple images with spectroscopic redshift are presently
available.  Figure~1 and Table~1 show this list of multiple images
with their properties. A number of strong lensing models have been
discussed in the literature
\citep{kneib96,allen98,ellis01,natarajan02} using only part
of the current available constraints.

Comparisons of the inferred lensing mass with the mass distribution
derived from the X-ray emission of the cluster pointed to a strong
discrepancy between the two estimators
\citep{miralda95,allen98}, if one assumes hydrostatic
equilibrium. A recent and detailed study of the gas distribution in
this cluster using high quality Chandra observations
\citep{machacek02} partly confirms this discrepancy, especially near the
center. It demonstrates that A\,2218 is probably not fully
relaxed. However a much better agreement is found in the outer parts
of the cluster where the X-ray mass compares with weak lensing masses
\citep{squires96,allen98}. Our analysis is however independent of the
physical state of the intra-cluster gas.

\begin{figure*}
   \centering \includegraphics[width=0.8\textwidth]{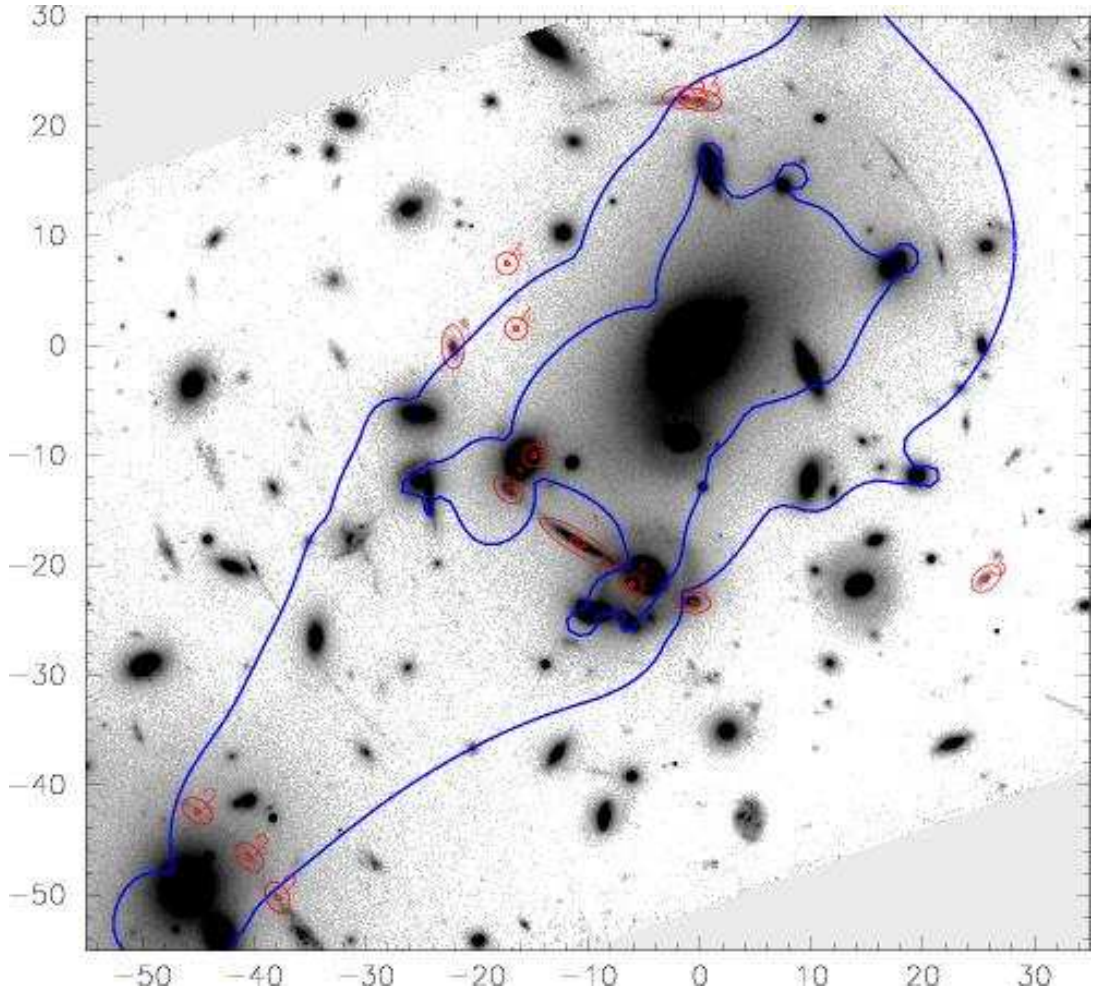}
   \caption{ Central part of the WFPC2 image of Abell 2218 displaying
   the 4 systems of multiple images (\#1 to \#4). The critical lines
   at $z_{\rm S1} = 0.702$ and $z_{\rm S4}= 5.576$ are indicated for
   the best mass model. North is up, East is left.}
   \label{a2218image}
\end{figure*}

\begin{table}
\caption{
Redshift and properties of the multiple images systems in Abell
2218. $N_i$ is the number of images detected in the {\em HST} image
and used in the optimization (note that a third image is predicted for
the system \#4 but is too faint to be detected). $N_{{\rm C}i} = 2
(N_i -1) $ is the number of constraints introduced in the
optimization. References: (1) \citet{pello92}, (2) \citet{ebbels96},
(3) \citet{ellis01}.  }
\label{table_zlens}
\begin{center}
\begin{tabular}{ccccc}
\hline\hline\noalign{\smallskip}
Multiple image system & $z_S$ & $N_i$ & $N_{Ci}$ & References \\
\noalign{\smallskip}
\hline
\noalign{\smallskip}
\#1 & 0.702 & 4 & 6 & (1) \\
\#2 & 1.034 & 3 & 4 & (1) \\
\#3 & 2.515 & 3 & 4 & (2) \\
\#4 & 5.576 & 2 & 2 & (3) \\
\noalign{\smallskip}
\hline
\end{tabular}
\end{center}
\end{table}

\subsection{The different gravitational lens components}
We start from the model described in \citet{kneib96}: the cluster mass
is distributed within two halos of dark matter centered respectively
on the main cD galaxy and on the second brightest galaxy. The mass
profile of each halo is modeled with a so-called truncated PIEMD (
``Pseudo-Isothermal Elliptical Mass Distribution'',
\citealp{kassiola93}, \citealp{kneib96}) 
characterized by 7 parameters: 4 are geometrical (center $(X_0, Y_0)$,
ellipticity $\varepsilon$ and position angle $\theta$) and 3 describe
the mass profile (velocity dispersion $\sigma_0$, core radius $r_{\rm
c}$ and truncation radius $r_{\rm t}$). Note that our definition of
the ellipticity is $\varepsilon = (a^2 - b^2) / (a^2 + b^2)$.
The numerical simulations of many different lens configurations 
and their fits by several analytical mass distributions have shown that
cosmological constraints are  not very sensitive to the choice of the
analytic model: provided there is enough freedom in the number of
parameters, the fits can easily adjust the true mass profile at the
location of the multiple images. For illustration if a characteristic
core radius is included among the free parameters, it will shrink to
very small values if the true profile is singular (see some examples in
the simulations of \citealp{golse02a}).  

Furthermore, we include the contribution of the 37 brightest cluster
galaxies with magnitude $m<19.5$ ({\em i.e.} m$_*$+2) and we
associated a truncated PIEMD to each of them. The geometrical
parameters (center, ellipticity, position angle) are fixed to those of
galaxy light parameters, and the mass profile parameters are scaled
with the total magnitude, using the prescription proposed by
\citet{natarajan97} and inspired from the standard Faber-Jackson 
and Kormendy relations:
\[ \sigma_0 = \sigma_{0*} \left({L \over L_*} \right)^{1/4} \quad
\theta_{\rm c} = \theta_{\rm c*} \left({L \over L_*} \right)^{1/2} \quad
\theta_{\rm t} = \theta_{\rm t*} \left({L \over L_*} \right)^\alpha
\]

\noindent $\sigma_{0*}$, $\theta_{\rm c*}$ and $ \theta_{\rm t*}$ are
reference values for a $m_*=17.5$ galaxy.  For the last relation,
$\alpha = 0.5$ means that a constant $M/L$ ratio applies to all
galaxies, while for $\alpha \neq 0.5$, $M/L$ scales as $L^{\alpha -
1/2}$. To minimize the number of parameters in the model, we fix
$\theta_{c*}$ to a very small value $\theta_{\rm c*} = 0.048''$ (or a
physical scale $r_{\rm c*} = 0.1 h^{-1}$ kpc) for a nearly singular mass
profile. Indeed, galaxies are essentially characterized by their
central velocity dispersion and the extension of their halo
\citep{natarajan97}.

\subsection{Optimization of the main lens parameters}
In order to avoid biases in the determination of the cosmological
parameters, we use the following procedure, already tested and
evaluated by previous numerical simulations \citep{golse02a}: using a
sparse sampling in the $(\Omega_{\rm M},\Lambda)$ plane (with steps of
0.1 for both parameters), we optimize all the other model parameters
with a Monte-Carlo method. Thus we do not bias the final optimization
toward a given cosmology. For each of the 2 main dark matter halos,
we adjust the 7 parameters of the PIEMD and for the individual
galaxies we explore the 3 parameters: $\sigma_{0*}, \theta_{\rm t*}$ and
$\alpha$.

A wide range of values is allowed for each parameter during the
Monte-Carlo initialization. When scanning the area $[0<\Omega_{\rm
M}<1,0<\Omega_\Lambda<1]$, the minimum $\chi^2$ is found for
$\Omega_{M}=0.001$ and $\Omega_\Lambda=0.9$, with a reduced
$\chi^2_{\rm min}=6.05$.  Similarly, assuming a flat Universe and
scanning the area $[0<\Omega_{\rm M}<1;-1<w<0]$ the minimum is located
at $\Omega_{M}=0.101$ and $w=-1$, with a reduced $\chi^2_{\rm
min}=6.65$. These high $\chi^2_{\rm min}$ values are likely
representative of yet non-perfect mass models and possibly also of
underestimation of intrinsic errors, especially in the image
positions. Anyhow, in both cases, the values of the parameters which
describe the two halo potentials and the galaxies are close to those
obtained by
\citet{kneib96} and \citet{natarajan02}. The main halo is centered on 
the central cD galaxy with a shift of a few arc seconds with respect to
it. The second halo is well positioned on the second brightest galaxy,
again with a few arc seconds shift with respect to its center
(Fig.~\ref{a2218image}).

The orientations of the two main halos are slightly different from those
of the light distribution but they show a clear tendency of alignment
between two halos. Moreover, the ellipticity of the main halo ($\epsilon_1
\simeq 0.28$) is significantly smaller than that of the cD isophotes
($\epsilon_{\rm cD} = 0.563$) while the ellipticity of the secondary
halo is quite large ($\epsilon_2 \simeq 0.61$). These features were
already pointed out by \citet{kneib96} in their lens model. They may
represent the signature of a merging phase of the secondary halo the
main mass concentration, in good agreement with the recent X-ray
analysis of the {\it Chandra} data by \citet{machacek02}.

The characteristic values of galaxy halo mass distribution are
compatible with current values deduced from galaxy-galaxy lensing
analysis like those found by
\citet{natarajan02b} in their study of 6 clusters of galaxies.  The
$\alpha$ parameter which represents the variation of the truncature
radius with luminosity is about 0.9. The value 0.5 which corresponds to a
constant $M/L$ value for all galaxies is ruled out, as already claimed by
\citet{natarajan02} and the $M/L$ ratio then scales as $L^{0.4}$. The
brightest galaxies seem to have a more extended and massive halo than
fainter ones, as already suggested in studies of the fundamental plane
of elliptical galaxies \citep{dressler87,jorgensen96}.

\subsection{Cosmological constraints}
Starting from the lens model determined in the previous section, we
perform a second level of optimization within the $(\Omega_{\rm
M},\Omega_\Lambda)$ or the $(\Omega_{\rm M},w)$ plane, with steps of
0.01. Only the most critical parameters are optimized in this
modeling, the others being fixed at their previously determined ``best
value". The fitted parameters are the velocity dispersions
$\sigma_{01}$ and $\sigma_{02}$, and the core radii $\theta_{\rm c1}$
and $\theta_{\rm c2}$ of the two main halos.  When changing the
cosmology, we keep the lens efficiency of the other clumps identical
by rescaling the velocity dispersion of the galaxies so that $\sigma_0
\propto (\sigma_{01} + \sigma_{02})$, therefore reducing the number of
free parameters.

\begin{figure}
   \centering
   \includegraphics[width=0.5\textwidth]{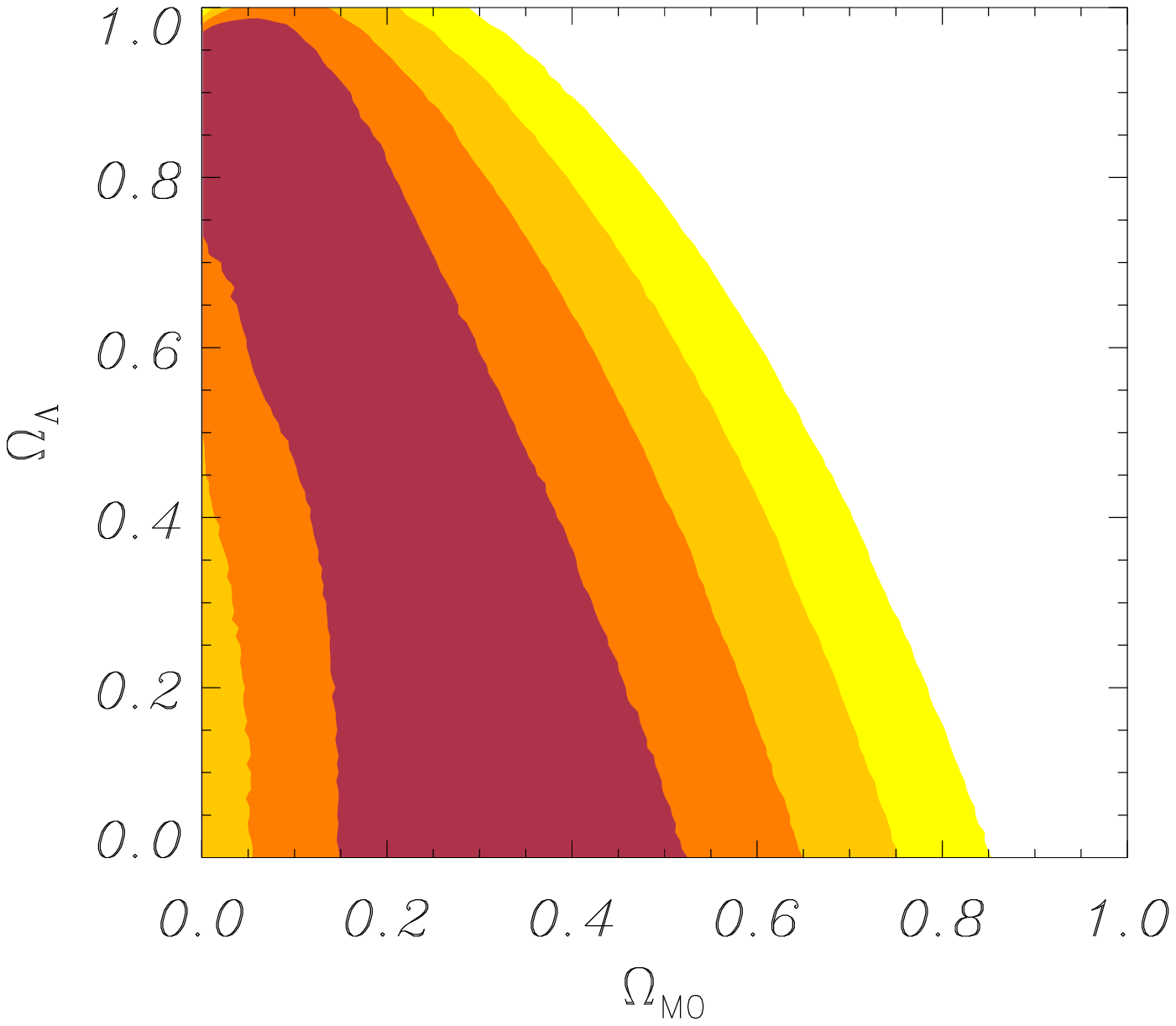} \\
   \includegraphics[width=0.5\textwidth]{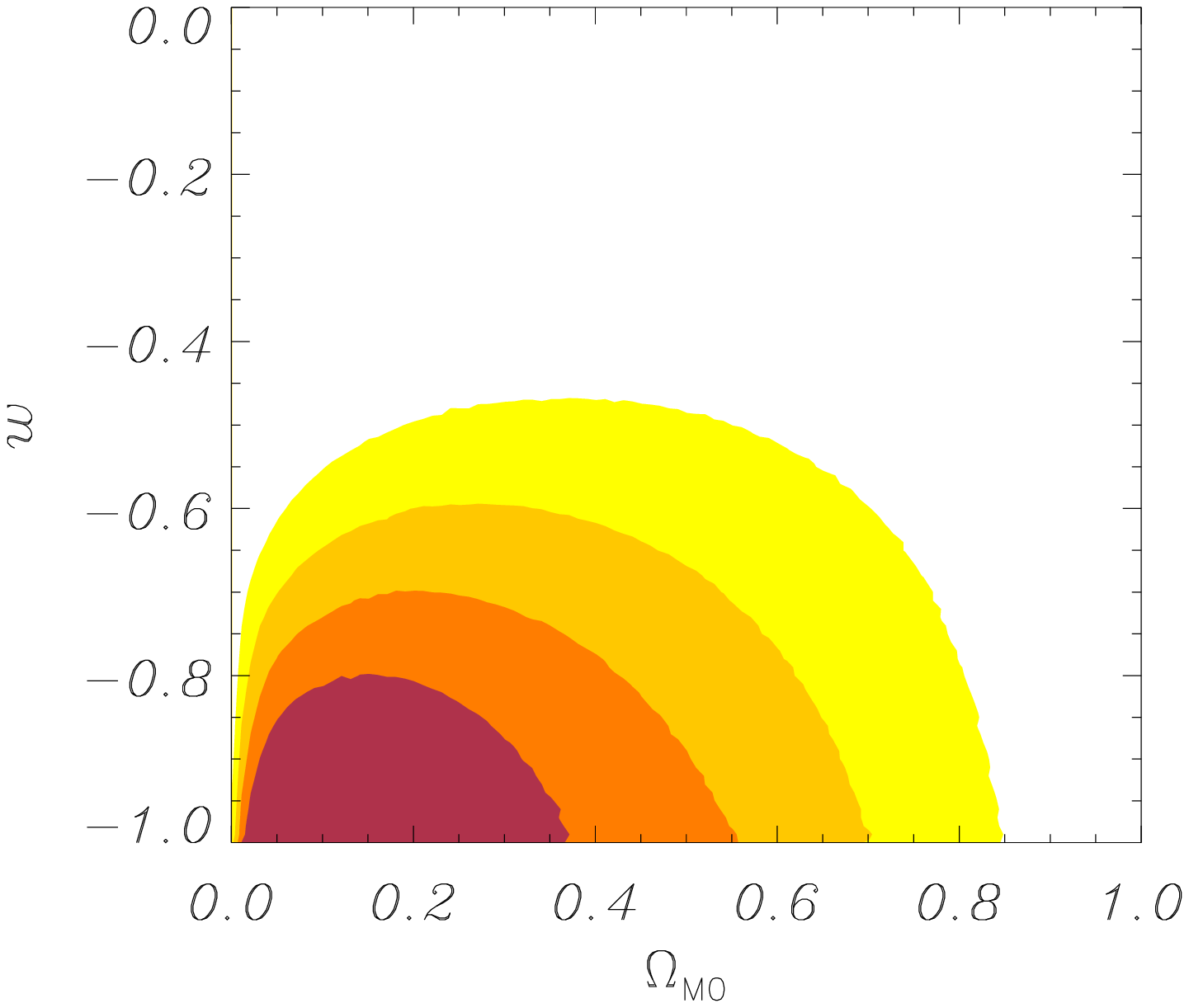}
     \caption{Confidence levels obtained in the $(\Omega_{\rm
     M},\Omega_\Lambda)$ plane (top) and in the $(\Omega_{\rm M},w)$
     plane (bottom) for a chi-square computed with $\nu = 10$ degrees
     of freedom (see text for details). Contour levels vary from
     $1\sigma$ to $4\sigma$ from the darkest to the lightest levels,
     {\it i.e.} probability levels are 68.3\%, 95.5\%, 99.7\%,and
     99.99\% respectively.}
   \label{chi2}
\end{figure}

The results of this optimization are displayed in
Fig.~\ref{chi2}. Only the best model and its $\chi^2$ value are kept
at each step, with no real marginalization on the lens
parameters. This is an approximation which is justified by analogy
with the analysis of CMB data: in the case of Gaussian distributions
of the errors, \citet{tegmark00} demonstrate the equivalence between a
full multidimensional marginalization and a much simpler
maximization. In our case, it is not clear whether the probability
distribution function of the lens parameters is close to Gaussian, but
to save computer time, we decided to use a similar approach. Further
analysis of the full likelihood distribution is necessary to fully
validate the method. But is is out of the scope of this paper,
presented mainly as a demonstration case applied on a single cluster
lens. In Fig.~\ref{chi2}, the confidence levels are given for a
number of degrees of freedom $\nu = N_{\rm C} - N_{\rm L} = 16 -
(4+2)=10$, where $N_{\rm C}$ is the number of constraints displayed in
Table \ref{table_zlens} and $N_{\rm L}$ is the number of parameters in
the optimization (4 for the cluster model and the 2 cosmological
parameters). The contours of these confidence levels follow very
clearly the expected degeneracy except at low $\Omega_\Lambda$ (see
\citet{golse02a} for a detailed discussion about these
degeneracies). Therefore we are confident {\em a posteriori} that the
modeling of the cluster is a fair representation of the true potential
and mass distribution.

\begin{table}
\caption{
Estimates of the uncertainty in the lens parameters derived from the
$1\sigma$ $\chi^2$ contours in the cosmological parameters space.}
\label{mod_errors}
\begin{center}
\begin{tabular}{ccccc}
\hline\hline\noalign{\smallskip}
& $\sigma_{01}$ (km s$^{-1}$) & $\sigma_{02}$ (km s$^{-1}$) & 
$\theta_{\rm c1}$ (\arcsec) & $\theta_{\rm c2}$ (\arcsec) \\
\noalign{\smallskip}
\hline
\noalign{\smallskip}
$(\Omega_{\rm M}, \Omega_\Lambda) $ & $1039^{+76}_{-20}$ & $383^{+15}_{-29}$ &
$18.60^{+1.30}_{-0.60}$ & $ 8.05^{+0.60}_{-1.88}$ \\
$(\Omega_{\rm M}, w)$ & $1035^{+19}_{-14}$ & $382^{+25}_{-2}$ &
$18.60^{+0.10}_{-0.60}$ & $ 7.93^{+1.01}_{-0.06}$ \\
\noalign{\smallskip}
\hline
\end{tabular}
\end{center}
\end{table}
In addition, if we overlay on the $1\sigma$ $\chi^2$ contours
the contours of the fitted lens parameters $(\sigma_{01},
\sigma_{02}, \theta_{\rm c1}, \theta_{\rm c2})$, we derive an estimate of the
variation of these parameters for the "best models". Table~2 summarizes
these values considered as an estimate of the error bars of the lens
parameters.

The consequences on the constraints of the cosmological parameters are
encouraging. It is clear from this analysis that an Einstein de Sitter
Universe is excluded, at more than 5\,$\sigma$. And as expected, the
constraints are more stringent on $\Omega_{\rm M}$ than on
$\Omega_\Lambda$. If we assume a flat Universe we get some narrow
windows on the parameters: \[
\Omega_{\rm M} <0.22 \quad {\rm or} \quad \Omega_\Lambda > 0.78 \] 
in close agreement with the constraints derived from Supernovae
experiments. And if we assume the existence of a dark energy component
in a flat Universe, we find
\[ \Omega_{\rm M}<0.37 \quad {\rm and}
\quad w < -0.80 \] These results are comparable to those
obtained by combining CMB and supernovae data ($w<-0.78$,
\citet{spergel03}).
Our results also compare well with other recent determinations
issued from the statistics of gravitational lenses.  For example, the
CLASS (Cosmic Lens All Sky Survey) survey of radio galaxies provided
constraints on $\Omega_{\rm M}$ and $\Omega_\Lambda$ or $w$ very
similar to ours, but with larger error bars \citep{chae02}.

However some limitations in the procedure are quite obvious. Sources
of uncertainties are shared between uncertainties in the image
positions (even with the accuracy of {\em Hubble Space Telescope
(HST)} images), and errors in the mass models. Indeed, although the
dependence in the mass profile has been addressed in our previous
paper, with accompanying simulations, the reality of the mass
distribution in clusters of galaxies is likely to be more complex
\citep{sand03}. Further examination of these limitations are in
progress and will be presented in a forthcoming dedicated paper.

\section{Conclusion and future prospects} 
We have shown in this paper that we can derive reasonable cosmological
constraints from the very detailed analysis of the lensing
configuration of the cluster of galaxies A2218. The necessary
conditions for this study are {\it simple}: deep multicolor {\em HST}
images of a well selected cluster-lens, identification of a minimum of
4 families of multiple-images systems and secure redshift measurement
of each family, which ranges from $z=$0.702 to 5.576\,. With these
constraints, and provided the mass distribution can be modeled by the
sum of a dominant component and smaller additional ones (all following
a truncated PIEMD mass profile), the geometrical problem can then be
solved. The cosmological constraints presented in this paper are of
similar accuracy than those derived from Supernovae
analysis. Interestingly, both analysis are purely geometrical and
completely independent tests, but they are not sensitive to the same
combination of distances thus providing nearly orthogonal constraints
in the $(\Omega_{\rm M}, \Omega_\Lambda)$ plane.

This new method to constrain the cosmological parameters is very
attractive, especially in view of the outstanding performances of the
Advanced Camera for Surveys (ACS) on board of {\em HST} and the
development of Integral Field spectrography that allow to secure the
redshift of many multiple images in a very efficient way. The very
spectacular ACS images presented by \citet{benitez02} on Abell 1689
show that in a very near future, we can use the proposed method as a
very serious cosmological test, by focusing on those clusters with
more than 4 multiple images with spectroscopic redshift. One advantage
of this method is its relatively low-cost in terms of telescope time
and relatively easy to implement - although progress is needed to
thoroughly explore the parameter space of the mass models and to
implement a fully comprehensive likelihood analysis. Such improvements
are currently under investigation and will in a near future allow a
better treatment of this exciting problem.

\begin{acknowledgements}
      We wish to thank B. Fort, R. Pain, M. Douspis and R. Blandford
      for encouragements and fruitful discussions. Part of this work
      was supported by the European Network {\em LENSNET:}
      ``Gravitational Lensing : New Constraints on Cosmology and the
      Distribution of Dark Matter'' of the European Commission under
      contract No : ERBFMRX-CT98-0172 and by the Programme National de
      Cosmologie of the CNRS. JPK acknowledges support from CNRS and
      Caltech. JPK and GS thank W.M. Keck Observatory and CFHT
      respectively, for their hospitality.
\end{acknowledgements}

\bibliographystyle{aa}
%\bibliography{referencesv2}

\end{document}